\documentclass[11pt, oneside]{article}   	
\usepackage{graphicx}				
\usepackage{amssymb}
\usepackage{amsmath}
\usepackage{endnotes}
\usepackage{bm}
\usepackage[T1]{fontenc}

\let\footnote=\endnote

\title{Demystifying AI in Criminal Justice\thanks{This paper is a draft of a chapter that will appear in the \textit{Oxford Handbook of AI and Criminal Justice} edited by Brandon Garrett.}}
\author{Richard Berk \\ 
University of Pennsylvania\\
\textit{berkr@sas.upenn.edu}}

\begin{document}
\maketitle

\section{Introduction}

In criminal justice settings, artificial intelligence (AI) is an outgrowth of conventional information technology (IT). For several decades, criminal justice organizations have been using IT for administrative tasks such as payroll, crime mapping as in COMPSTAT, sentencing guidelines, inmate risk assessments, and much more. These are either fully or partly ``algorithmic'' and in today's world can be relabeled as AI. 

AI itself has absorbed statistical advances going back at least to the 19th century (Stigler, 1990; 2002). One example is ubiquitous linear regression. In the early 19th century, Adrien-Marie Legendre and Carl Friedrich Gauss independently developed the method of least squares, which is the mathematical foundation of regression analysis and much more. Gauss also  showed how least squares is connected to maximum likelihood estimation. In the late 19th century and early 20th century, Karl Pearson established the matrix algebra framework for regression and correlation and extended the theory to multiple variables. By the middle of the the early 20th century, R.A. Fisher completed the formal integration of regression into the theory of estimation, hypothesis testing, and experimental design and in so doing, clarified the difference between correlation and causation. There are rich histories also for principal components analysis, factor analysis, logistic regression, classification and regression trees, and many basic concepts and procedures from statistics. A lot is new in AI, but a lot is just a repackaging of existing statistical tools. 

If one takes the media seriously, AI applications also are novel. Here too, however, there there is a very long history. The earliest scientific weather forecast is said to have begun with Aristotle's "Meteorologica" (Goldstein, 2002). A precedent-setting form of industrial quality control emerged in England in the 13th century with the testing of coins to determine if the English mint was meeting its required standards (Stigler, 2002). In the late 1600s, Lloyds of London was doing risk assessment to set premiums for the ocean transport of commercial goods (Palmer, 2007). Machine oversight in the form of thermostat control is credited to Cornelis Drebble in approximately 1620 (Tierie, 1932). In the late 19th century, forensic science was significantly advanced with Francis Galton's the use of pattern recognition for finger print identification (Stigler, 2002). Educational testing for school admission and placement was put on more rigorous foundations with the Binet-Simon IQ Test in the early 20th century (Binet and Simon, 1907). The first autopilot was developed by the Sperry Corporation in 1912 based on the work of Lawrence Sperry. Autonomous weapons are old news as well. Once placed in the ground, land mines can be triggered when they sense pressure or vibrations. Humans are not in this loop. They have been used by the U.S. military, its allies, and its adversaries at least since World War II.

Yet, because of the recent computer processing power gains, very large data sets that now can be downloaded, scraped, or compiled, and of hugely clever computer code enabling the two to interact, a variety of very creative applications have materialized. Criminal justice applications are no exception. There are new IT procedures, and the reach of many existing tools has been extended. Like the familiar metaphor of a frog in a kettle of slowly warming water, it is important to consider whether these gradual IT changes have placed important criminal justice activities at a moral or functional tipping point. And the threats go far beyond criminal justice. Even Pope Leo XIV has entered the debates (Stancati et al., 2025).

The material that follows is written for readers with little or no background in statistics or computer science. It is not intended to replace more technical treatments. It is intended to supplement them and encourage readers to dig more deeply into topics that strike their fancy. Several textbooks are listed in a long footnote near the end of the chapter, but they all assume some background in statistics and/or computer science. I could find nothing that was less technically demanding that was also technically sound. 

There also are many good treatments on the internet, some as documents that can be downloaded and some as videos that can be watched. In both cases, however, caution is needed. There is no peer review, and some are commercially motivated. A good way to start is with the few that have existed for at least a several years and for which the informed reviews of generally positive (e.g., Kahn Academy, 3Blue1Brown, and treatments associated with major universities such as MIT, Harvard, and Stanford). 

\section{What Is AI?}

AI is a complex amalgamation of some things old and some things new. The ``I''  for ``intelligence'' introduces a metaphor that has become the stuff of criminal justice science fiction (e.g., Blade Runner, Minority Report, Robocop). Vendors of criminal justice software commonly market their products as artificial intelligence (e.g., TRULEO Inc.). Criminal justice and law journals are currently larded with commentary on AI, sometimes badly informed or at least badly dated. Examples include concerns about bias in AI facial recognition (Jones, 2021) and in criminal justice risk assessment (Farayola et al., 2023). More clarity about AI definitions and mechanisms seems like a good place to start (Bellovin, 2025).  

\subsection{Defining AI}

There is no consensus definition of artificial intelligence, even among those most familiar with  its technical details (Saghiri et al., 2022). The following definitions of 
AI provide a sense of its span.
\begin{itemize}
\item
The U.S. National Artificial Intelligence Act of 2020 defined AI as  ``A machine-based system that can, for a given set of human-defined objectives, make predictions, recommendations or decisions influencing real or virtual environments. Artificial intelligence systems use machine and human-based inputs to: (A) perceive real and virtual environments; (B) abstract such perceptions into models through analysis in an automated manner; and (C) use model inference to formulate options for information or action.'' 
\item
The EU defined AI far more briefly and narrowly as: ``Systems that display intelligent behavior by analyzing their environment and taking actions, with some degree of autonomy, to achieve specific goals.''\footnote
{
The U.S. and EU definitions are part of a draft report from the Advisory Committee on Artificial Intelligence Impact and Potential in Pennsylvania. The author of this chapter is a member of that committee.
}
\item
Alan Turing's famous definition builds on machines or software that can perform tasks typically requiring human intelligence. Related definitions focus on AI procedures that simulate how the human mind works. Some argue that we have passed that threshold.
\item
Perhaps the broadest definition is any AI procedure that uses statistical techniques for large datasets.
\item
Probably the most sensible AI definition that is useful in criminal justice practice is narrow and resurrects features of the Turing formulation: \emph{AI is an algorithm able to perform a criminal justice task that ordinarily requires human intelligence.} But some caution is needed.  Human intelligence is not defined and comes in many variants and skill levels. Moreover, existing definitions, even when a product of sophisticated thinkers, use concepts whose meanings are unclear as well (Neisser et al., 1996).
 
\end{itemize}

In contrast, there is widespread agreement that there are several AI developmental stages. A common hierarchy has three levels.

\begin{enumerate} 
\item
At the lowest level, ``Narrow AI'' performs a single task that is well defined and is often superior to humans doing that same task. Recidivism forecasts for prospective parolees are an example (Berk. 2017). Currently, virtually all AI is narrow. 
\item
At the next level is ``General AI,'' which can move across many tasks and domains without being trained on each. For example, a police officer performance evaluation tool might effectively forecast which police officers are likely to use unnecessary force, which police officers are likely call in sick, and which police officers are likely to perform poorly on the paper and pencil tests necessary for promotion. Moreover, the procedure performs well in urban and rural jurisdictions alike.  At this point, many universities and technology firms are actively working on general AI. Transfer learning is a step in that direction (Weiss et al., 2016; Yang et al., 2020). In transfer learning, AI procedures trained in one domain, learn how transport their methods to new domains. The computer science term is ``fine tuning'' (Vrban\v{c}i\v{c} and Podgorelec, 2020).
\item
At the highest level is ``Super AI,'' which is not well defined and even might not be attainable (Johnson, 2025). Super AI refers to cognitive skill that surpasses humans on almost anything and might have self-awareness and emotions. Super intelligent algorithms might set goals independently of humans and be able to re-program themselves to achieve those goals. Such capabilities are the stuff of science fiction and the fodder for claims that AI poses an existential threat to humans (Fitzpatrick, 2025). Yet, it remains the holy grail for many (Stokel-Walker, 2025).  

\end{enumerate}

\subsection{Algorithms Not Models}

AI is powered by algorithms, and there are important differences between algorithms and models (Breiman, 2001). ``At its most fundamental level, an algorithm is nothing more than very precisely specified instructions for performing some concrete task'' (Kearns and Roth, 2020: page 4). A recipe for sourdough bread is an algorithm. Steps required for filing income tax forms is an algorithm. The open source code for Meta's large language model (LLM), called LLaMA, actually is a very complicated algorithm despite its  name. 

A model is a way to characterize some phenomenon. Its goal is explanation leading to understanding. Newton's second law is a model: Force = Mass X Acceleration. It follows that a model can be right or wrong. An algorithm cannot. Either an algorithm performs in a satisfactory manner or it does not. There is nothing to be right or wrong about. 

Discussions of AI can run off the tracks when models and algorithms are confused or conflated. Often this starts with sloppy language. But, the issues can get subtle. For example, what does one call computer code built with a conscious effort to incorporate certain features of models in an algorithm. This is currently an active research area in physics and computer science (Hao et al., 2022). Imposed on an algorithm might be constraints coming directly out of thermodynamics (Brunton and Kutz, 2022: chapter 14).

\subsection{Robots}

Robots come in many forms designed to do many different tasks. Some robots work on assembly lines; some, renamed as chatbots, are widely used to provide information at websites where products or services are sold; some provide companionship to lonely humans; some are realized as self-driving automobiles or semi-trucks; some are autonomous drones; some do surgery on humans, and so on. Robots give AI the ability to directly act on its environment and become, in effect, ``embodied'' AI. 

There are already in service robotic police dogs (``Spot'' https:// bostondynamics.com/) and robotic police officers (``K9'' https://www.nytimes.com/ 2023/10/03/learning/police-robot.html). Robodogs have been used for search and rescue after natural disasters, to intervene in hostage standoffs, and provide intelligence to guide police officers before they enter a dark building.  Robocops have been used for routine surveillance on subway platforms, making traffic stops that could be dangerous for human police officers, de-escalating dangerous situations (e.g., delivering a cellphone to a barricaded and armed offender), and even applying fatal force (https://builtin.com/  robotics/police-robot-law-enforcement).

Robots combine hardware and software in a collaboration that will continue to improve. When currently a robot does not perform as expected, flawed AI often is blamed. But the problem might be found in the hardware instead. Going forward, however, the difference between hardware and software will be less clear as ``AI on a chip'' becomes increasingly widespread. This technology integrates artificial intelligence capabilities directly into a computer chip so that AI tasks, such as image recognition, video processing, and speech manipulation can be performed locally, without relying on external servers or cloud computing. Improvements in speed, energy  efficiency and privacy can follow (Viswanathan, 2020; geeksforgeeks, 2024).\footnote
{
Further obscuring the distinction between hardware and software is very recent work that skips the need for either and builds the beginnings of AI capabilities directly into electric circuits (Dillavou et al., 2024). The work is quite remarkable.
}
For this discussion, the focus is on software only, but further blending of hardware and software should not materially affect the material presented.

\section{How Does AI Work?}

From a high altitude vantage point, AI rests on two broad processes. The first is data preparation. In the old days, that was called feature engineering. Farther back, it was called file construction. The second is searching in the prepared data. Combined, AI is much like an extraordinarily powerful browser that searches through a dataset once it has been prepared in the first process. Neither could be implemented in their current form without the enormous increases in computing capability that has unfolded over the past several decades. The search process is easier to explain than the data preparation. Searching is addressed first.

\subsection{AI As a Very Powerful Browser}

To understand what makes AI distinctive, it helps to begin with the contrast between closed-form and open-form solutions. Although ``open form solutions'' is not standard term, the contrast captures the nature of AI in an easily digested form. Traditional statistical models rely on closed-form solutions, which are fixed formulas derived from statistical theory.  AI systems operate far more flexibly. They search through vast spaces of possibilities, adjust to patterns in data, and adapt their structure over time. Their outputs are not the result of applying a fixed rule but of learning those rules from examples. In this sense, AI shifts from deduction to induction, from specification to discovery.

At first glance, a closed-form statistical solution might resemble a statute. Both offer fixed, rule-based procedures for producing an outcome. But just as statutes are often written in general terms and require interpretation in practice, closed-form prescriptions may not fully resolve ambiguity or accommodate variation in the real world. The meaning of a statute evolves through litigation that interprets, refines, and adapts the law based on the specifics of particular cases brought before a judge who interprets and expands upon the arguments offered by the litigators. In this sense, the legal system incorporates a kind of learning layer on top of a static foundation. What distinguishes AI is that the entire process, from absorbing examples to refining patterns to applying rules, is automated. It does not rely on human judgment to adapt. The system adjusts itself. A closed-form solution stops at the recipe. AI learns how to modify the recipe as new cases are presented.\footnote
{
Portions of the explanatory language and metaphor development for closed form solutions were assisted by OpenAI’s ChatGPT (accessed July 2025).
}

Consider a simple data-based illustration to help fix these ideas. Suppose there are data on 100,000 probation sentences for burglary given to convicted offenders in a particular jurisdiction over several years. One might wonder about the typical length of the probationary period because the supervisory staff needed must be anticipated. A human then should decide what is meant by ``typical.'' The ``average'' length of probation sentence might be selected. How might a computer determine that value through search?

Suppose there exists a formula for the mean or median that can be applied to compute an average. This is called a closed form solution that for 100,000 cases would be too onerous for a human to apply, but is almost instantaneous for a computer. The computer is serving as a powerful pocket calculator.  Even though a computer can compute the mean or median of 100,000 probation sentences far faster and more accurately than humans, closed form processes are not usually called AI. 

Suppose there is no such formula, but a human in charge knows what properties the measure of  ``typical'' should have. For example, it might be the probation sentence length that is the closest overall to each of the 100,000 cases. A single closeness number is sought representing for all the cases at once a central tendency. 

``Closeness'' can be defined as the distance of each of the 100,000 probation sentence lengths from any prospective numerical value for a sentence length, central tendency. One approach might be to use the arithmetic difference in sentence length between that prospective number and the length each probationer's supervisory sentence. For example, if a possible value for typical is 8 months and a given probation sentence is 12 months, the closeness is 4 months. 

By that reasoning, however, there will be lots of closeness values that are negative;  a prospective typicality measure minus the value of a sentence length will negative if an actual sentence length is longer than a prospective measure of overall closeness. Negative closeness values, however, may be unintuitive, especially when thinking about distance. There are two obvious solutions: for a closeness metric, the absolute value each difference could be used or the square of each difference could be used.  This ensures that all closeness values are zero or positive. 

With either of these definitions, a sensible aggregate, typicality value might be found for which the sum of difference values is as small as possible. Such a number would place the typicality value as close as possible in the aggregate to the 100,000 probation sentencing values. It would qualify as a measure of typicality. Moreover, if the absolute value definition of a difference is used, the median sentence length will be estimated. If the squared value definition of a difference is used, the mean sentence length will be estimated.

It is then time to put the computer to work. A very large number values for ``typicalness" can be tried (e.g. millions), and the selected value would be the one that maximizes aggregate closeness by minimizing the sum of distances. This is an impractical task for a human, but with efficient code is trivial, even on a laptop. This ensures that all closeness values are zero or positive. One might say that the computer \emph{``learned''} what typicality value is the best typicality value for the distance definition being used. In today's world, that would probability be called AI, and in particular \emph{machine learning}. This nudges us into the realm of anthropomorphic metaphors. Algorithms are said to ``learn'' or ``discover'' patterns.

Underlying these searches is a formal objective, typically expressed as an objective function. In this example, the search has a criterion for success that defines an objective: minimizing the sum of the distances, whether their absolute values or their squared values. The technical term is an ``objective function,'' sometimes called a ``loss function'' or ``cost function.'' Formally, an objective function is a mathematical expression that defines what needs to be minimized or maximized in an optimization problem. It quantifies the goals of the procedure and often has a procedure-specific name. For example, minimizing the sum of the squared differences just described is called ``least squares,'' which is widely used in a variety of statistical procedures often leading to closed form solutions. 

Although all AI search procedures have an objective function, some do not require that the function be optimized. Sometimes, performing ``well enough'' compared to the available alternatives will suffice.  In management science, this is often called  ``satisficing'' (Simon, 1956).  An example, discussed briefly later is reinforcement learning. 

Searching for the mean or median generalizes in many ways. At some point as the searches become more complicated, searching  by ``brute force'' starts matter. A search is no longer even close to instantaneous. For example, one might care about how alike each pair of the 100,000 sentences happen to be. A histogram of these pairwise disparities might motivate a discussion of sentencing disparities across cases.

The appropriate combinatoric formula is,

\[
\binom{n}{2} = \frac{n(n - 1)}{2},
\]
where in this example $n=100,000$. The result is 4,000,950,000 comparisons perhaps using the squared difference between sentence lengths for each pair of probationers as a measure of dissimilarity. 

There is now a need for computational shortcuts while the algorithm still, in effect,  searching through data. AI now gets very mathematically sophisticated in how it ``learns'' the relevant characteristics of the data. Many approaches help guide the search over passes through the data such that the quality of the results generally improves with each pass. Perhaps the archetype is gradient descent (Goodfellow et al., 2016: section 4.3).\footnote
{
Imagine water running down an irregular hill.  To get to the bottom, it tacks left and right, each time picking the path that is the steepest downhill route. This idea can be applied to a search algorithm to improve search efficiency. If the solution being sought is at the bottom of an objective function valley, proceeding like flowing water down a real hill is an excellent strategy. But it can fail if the water reaches a choice point where in the immediate vicinity all of the available routes lead upward rather than downward. The water reacts as if it has gotten to a valley bottom where all path are up. But this stopping point may not even be close, and is called a ``local solution.''
}
Very soon, sophisticated shortcuts start to dominate the search process. These commonly depend on the kind of data on which a search is undertaken. We are ready now to consider data preparation.

\subsection{How The Data Type Affects The Search}

In a first statistics course, one learns how different kinds of variables affect the choice of statistical procedure. To take a simple example, if a variable is numeric (e.g., age in years) the central tendency measure can be the mean, mode or median. If the variable is categorical (e.g., employed or not employed) the central tendency measure is computed with proportions. For instance, if out of the 100,000 probationers .30 were employed when first arrested, and .70 were unemployed when first arrested, unemployment is more typical than being employed. These kinds of differences affect the AI search. 

For AI, the \emph{kind of dataset} itself becomes very important. Data that arrive in a csv file ready for an Excel spread sheet makes things very easy. There is decades of craft lore on how to properly proceed. But what does one do with facial images that might be the raw input information to a facial recognition algorithm? What about text from a court a transcript being used to determine if testimony from a prosecution expert witness should have been disregarded as not meeting the Daubert standard? What about speech recorded from 911 calls? Let's consider facial recognition first because many readers are familiar with at least some of the issues. The discussion must get into the weeds a bit to appreciate better what can be novel about AI

\subsubsection{Images of Faces and Other Things}

Facial recognition AI is designed to match a face in one image (e.g., from a security camera) to a face in another image (e.g., a mug shot). How might similarity between faces be defined? One option would be to measure distances between notable facial features such as between the eyes or between the tip of the nose and top of the upper lip. There might be a dozen such metrics. One could compute the proportion of times over the set of such metrics that a particular facial image is within some margin of error of another facial image. The larger the proportion, the more likely the match.

This may seem easy, but it is very hard. For example, a face in different images may have different orientations. A face in one image might be vertical. The same face in a comparison image might be rotated clockwise about 30\textdegree.  Because computer code is very literal, a face rotated 30\textdegree  is not treated the same as the same face at 90\textdegree. Moreover, working from the modest number of facial features that humans might recognize is too crude. Indeed, it is one of the problems with eye witness testimony (Lane and Houston, 2021). 

There actually is a host of other visible features that could matter: smile wrinkles, pierced ears, a mole on the left cheek, sun damaged skin high on the forehead, a chipped front tooth that shows when smiling, a scar on the left side the chin, sleep deprivation circles under both eyes and so on. And even if all such facial features were listed, how would an algorithm identify them in an image? 

A far more abstract approach is needed that is less tied to human visual conventions. An empirical, \emph{facial deconstruction} can follow. The process is sketched over the next several pages, ideally to provide some intuitive understanding. There are a host of important details that are beyond the scope of this chapter.

Facial recognition algorithms commonly segment a facial image into ``patches" that are a bit like large pixels. It is as if a grid is overlaid. Patches, often called ``tokens,'' initially are examined for very simple patterns such as edges, corners or gradients. The overall intent is to gradually build up from the simple to the complicated. Consider horizontal forehead wrinkles. In a single patch, a wrinkle is just an edge, identifiable by vertical changes in skin color and brightness.
 
There can be communication and cooperation across patches. For example, suppose adjacent patches to the left and right find much the same thing.  These results can be shared among the three patches, suggesting a continuous horizontal edge across them.

Some distance away, five patches may have identified another but similar horizontal line perhaps produced by a different forehead wrinkle. The findings from both sets of patches are then shared in part because the two horizontal patterns look rather alike. In other words, each patch can be filtered for relatively simple patterns that may be spatially generalized. All of this occurs in what computer scientists often call the first ``layer.''\footnote
{
The term ``layer'' comes from the way the code is written and is introduced here because it is commonly used in expositions written by computer scientists and those who work with computer scientists.
}

In a second layer, the patterns initially found are assembled to create a new image - one that includes only the simple features discovered in the first layer. Other more complex facial characteristics are then extracted using another set of filters. The results from this later layer are less abstract and may correspond to familiar facial structures. For example, two dark ellipses toward the middle of the image may be located just below and centered on a vertical ridge. Perhaps an a rough approximation of a nose has been found. Sometimes the patches are resized as information is aggregated to improve detection while keeping the number of detected patterns manageable.

There usually are subsequently several more layers motivated in a similar manner: construct a new image from the findings of the preceding layer, and apply as needed another set of filters. Early layers are meant to detect very simple, rudimentary patterns. Middle layers combine these to identify shapes and facial parts. Late layers assemble complex patterns such as whether the eyelids droop. However, many of the patterns discovered are not interpretable or named in human terms.

The process stops when there are enough assembled features to describe a face well. The output from a last layer is treated as a high dimensional vector of predictor variables that were derived sequentially from the raw image. These variables then can be used to identify or verify particular individuals. 

For training algorithms, the identity of the image (called a ``label'') is known. A test of the algorithm is whether, when presented with a set of labeled facial images, it can correctly identify each of the ones on which it was trained. Features of the facial deconstructions are compared. Critically, these new predictor variables were not defined in advance by humans; they are learned from the image data. 

For real applications, such as training an algorithm to recognize faces from a lineup or from closed circuit TV images, the algorithm learns from thousands of facial images. Each is deconstructed as just described and each is labeled. 

Suppose there are 1,000 facial images, each deconstructed into 50 predictors. The assembled data could have 1,000 rows and 51 columns with one row for each facial image in the training data, one column for each predictor and one more column for the images' labels. The algorithm searches to find weights for each column such that that each row has a very high probability of finding its correct label. In statistical terms, this is a very large classification exercise. When those probabilities are sufficiently high, the facial recognition algorithm is declared trained; the algorithm can correctly identify a face much (or maybe most) of the time.

Once trained, facial recognition algorithms are deployed “in the field” where the true label is unknown. For example, a smartphone may capture images of several individuals just before a violent incident. If any of those individuals were part of the training data, there is a good chance the algorithm will identify them. That is one reason why very large and diverse datasets are used in training.\footnote
{
Large datasets come with large risks too (Petkauskas and Japienyte, 2025).
}
  
 A more common application is to have facial images with known labels that can be deconstructed by the algorithm (e.g., mug shots). There are also facial images with \emph{no labels} that can be deconstructed by that algorithm (e.g., a smart phone image of several individuals about to break into a jewelry store). The two sets of deconstructions can then be compared for possible matches. The more weighted, deconstructed predictor values they have in common (e.g., three forehead wrinkles rather two or four), the more likely there is correct match.

More generally, all images, whether facial or not, are processed in broadly similar ways. For example, the algorithms guiding autonomous automobiles learn from the deconstructed images to correctly identify trees, stop lights, baby carriages, stop signs, bicycles, telephone poles and other many other objects.  Gained is the ability subsequently to recognize those objects when there are no labels attached.\footnote
{
From a distance, processes just describe sounds a like routine statistical procedures used in forensics such as identifying counterfeit coins (Christian, 2023). But there are important differences. For example, the statistical procedures used to identify counterfeit coins cannot accept image data because their data preparation tools are far less elaborate, and the classification problem uses labels with relative few outcome classes (e.g., counterfeit or not). 
}

\subsubsection{Large Language Models as a Special Case of Generative AI}

Facial recognition algorithms and image recognition more generally are a small fraction of all AI applications in criminal justice. There are several more flavors of AI that can be discussed in this venue, but far more briefly.
 
\emph{Generative AI} is any artificial intelligence system designed to create new content. This content can take many forms such as:
\begin{itemize}
\item
Text (e.g., writing reports, poetry, or computer code)
\item 
Images (e.g., art, ``fake'' photos)
\item
Audio (e.g., music, ``fake'' speech)
\item
 Video (e.g., ``deepfakes,''  animations)
 \item
 3D Models (e.g., molecular structures, architectural mockups). 
 \end{itemize}
 Applications in criminal justice include:
 \begin{itemize}
 \item
 Drafting legal arguments (e.g., summaries of relevant statutes)
 \item
Writing a report from the video images (e.g., from a body worn camera)
 \item
Simulating cross-examinations (e.g., of expert witnesses)
\item
Conducting interviews (e.g., with eye-witnesses)
\item
Summarizing case law (e.g., for charging a hate crime)
\item
Translating languages (e.g., Spanish to English)	
\item
Writing computer code (e.g., for the analysis of financial data in a complicated tax evasion case)
\end{itemize}.

These applications are ``generative'' in the sense that their output is not from memorization or just retrieving existing information. New content is provided. Generative AI systems are often built using deep learning architectures such as ``transformers,'' ``autoencoders,'' and ``GAN''s (Generative Adversarial Networks), depending on the tasks undertaken. These are defined in the next section.

Large Language Models (LLMs), such as GPT or Claude, are a special case of generative AI trained not on images but on text. Yet many of the same underlying ideas apply. Just as a facial recognition, when the algorithm begins by segmenting a facial image into patches, an LLM starts by breaking a body of text into “tokens.” A token is usually a short sequence of letters, a whole word, or even a punctuation mark. For example, the sequence “breaking a body of text” could be tokenized into something like “breaking,” “a,” “body,” “of,” “text.” These tokens serve the same purpose as patches in image recognition: they are the atomic units from which more complex patterns are built.

Each token is first passed through filters to extract very simple information such as how often the token appears, its grammatical role, or whether it is similar in use to other tokens nearby. These filters are applied in the first layer of the model. Like image models that identify edges or corners early on, LLMs begin by finding elementary relationships between tokens. For instance, the word “court” may often appear near the word “judge” in a paragraph, sentence or phrase. These relationships are then passed to a later layers, where they are aggregated and refined. In middle layers, the model might learn that certain clusters of tokens form legal arguments or a have narrative structure. At even later layers, the overall tone of a paragraph might be characterized as adversarial or that a sequence of sentences contains a prosecution’s theory of the case. Just as with faces, many of these patterns are not labeled or interpretable in human terms, but are abstract combinations learned automatically from enormous volumes of text.

The final layers of the model produce a high-dimensional summary vector, which is a kind of abstract fingerprint of all the preceding tokens and their learned relationships. In generative mode, the LLM uses this fingerprint to predict what tokens are likely to come next, one at a time, thereby generating new sentences. During training, it compares its predictions to the actual next token and adjusts internal weights to improve accuracy. Like facial recognition systems that match a new image to a labeled mug shot, an LLM is trained to match a given prompt it gets to an appropriate continuation. And just as facial recognition depends on thousands of labeled images, LLMs are trained on billions of words from books, websites, transcripts, and legal decisions. When deployed, these models generate plausible and grammatically correct text even when the input is new and unlabeled. In short, an LLM is a patch-token-filter-layer system for language rather than for images.

\subsection{An AI Menagerie}

Professional reputations in statistics and computer science are enhanced with the discovery of novel procedures. The more novel the better. An appearance of novelty can be conveyed with clever procedure names even when the advance claimed is trivial or no advance at all. A battle over scientific turf can follow. The AI procedures discussed briefly in this section are named as I interpret the technical literature. Others may legitimately disagree.

Arguably the common denominator for all AI is the need to search for patterns in data. But how such searches are conducted can help define different flavors of AI. Data preparation of many different forms, in turn, helps to differentiate one AI flavor from another. Most fall under the broad umbrella of machine learning. 

\emph{Machine Learning} a branch of artificial intelligence in which computer algorithms improve their performance at a task through passes over the data, aiming to optimize some objective function. Gradient descent is a common approach. Machine learning, therefore, is a large umbrella under which many AI variants can fall. 
\begin{itemize}
\item
\emph{Supervised Learning} is a type of machine learning in which the algorithm is trained on labeled data. For such data, both the inputs and correct outputs are known for each training observation, or what computer scientists call an ``example.'' Supervised learning  can be seen as a form of conventional, but nonparametric, regression analysis.  That is, no functional forms are imposed on the search. The functional forms are discovered. The goal is to learn associations that map inputs to outputs, so the algorithm can correctly forecast outcomes for unlabeled cases. At arraignment, for instance, a risk tool will forecast which individuals are a flight risk. But risk assessments also can be backward looking in time when used for forensics. For example, a supervised learning algorithm can be trained to determine which characteristics of coins are associated with whether a coin is counterfeit. Once that is determined, the trained algorithm is able to resolve whether an unlabeled coin manufactured in the past is counterfeit (Christian, 2023).\footnote
{
Linear regression can be written as $Y =  \bm{\beta} \mathbf{X} + \epsilon$, where $Y$ is a response variable such as sentence length, $\mathbf{X}$ is a set of explanatory variables such as the crime of conviction and the number of prior arrest, $\bm{\beta}$ is a set of weights (one for each explanatory variable), and $\epsilon$  essentially is noise. There is a closed form solution, and each explanatory variable is related to the response variable as a straight line. In machine learning, $Y = f{X} + \epsilon$, where $f(\mathbf{X})$ is just some unspecified function instead of a linear one. Machine learning algorithms search for an estimate of that function. For one explanatory variable the function might be S-shaped, and for another explanatory variable the function might be U-shaped. 
}
\item
\emph{Unsupervised learning} is a machine learning approach in which an algorithm is given data without labeled outcomes and must discover patterns or structures using only the predictor variables. Common tasks include clustering similar items or reducing the complexity of the data. Suppose, for example, in a given jurisdiction cases are assigned at random to different judges. A clustering algorithm should be able to partition the data so that post conviction, judges who give longer prison sentences than 5 years on the average are one cluster, and judges who give average prison sentences of equal to or less than 5 years on the average are in another cluster.
\item
\emph{Reinforcement learning}  is a trial-and-error approach where an ``agent'' learns to make a sequence of decisions by taking actions in an environment to maximize a reward over time. When the ``best'' sequence of decisions is determined, it can provide guidance for policy decisions. For example, suppose there are data from body worn cameras on large number of police encounters with individuals who from the 911 calls for service are likely to be mentally ill or suffering from dementia. There are several things an arriving officer could sequentially do. But which should be done and in what order? Reinforcement learning might be able to determine from the videos which sequence of actions by a police officer is least likely to lead to the use of force. Although reinforcement learning has an objective function, the usual goal is satisficing not optimizing. For some, that means that reinforcement learning is not machine learning.
\item
\emph{Transfer learning} is a machine learning method in which knowledge gained from working successfully on one task is reused to help undertake a related but different task, often with less data. It is especially useful when labeled data for the new task is scarce.  For example, a risk assessment algorithm trained on large jurisdiction with relatively many arraignments per week (e.g., Philadelphia, Pennsylvania) might need to be used in a small jurisdiction with relatively few arraignments per week (Chester, Pennsylvania). The intent would be to leverage a risk tool trained on a large number of examples to improve the algorithm's performance in a jurisdiction having many similarities to the larger jurisdiction, but on a far smaller scale. 
\item
A \textit{neural network}, a metaphor for how the neural structure of a human brain works, is a statistical method used to search in data for highly nonlinear relationships. It can be a form of machine learning depending on how it is used.  Algorithms based on neural networks, or used in concert with neural networks, include:
\begin{itemize}
\item
\textit{Deep learning} is a richer form of neural networks containing ``hidden layers.'' Hidden layers are landing spots in which intermediate results are stored. They are hidden in the sense that they make no direct use of the raw data. They are  a computational device to aggregate earlier results, make them more complex, and then to use those results as inputs to the next hidden layer, much like the role of layers in facial recognition algorithms. Deep learning is commonly seen as a form of machine learning.
\item
A \textit{transformer} is a deep learning architectural feature that processes data by letting each element (such as a word or image patch) capitalize on associations with all others in the input. It can capture relationships regardless of their location in the image or text. The goal is to find structure in a series of tokens. It is typically applied as data preparation before a deep neural network is employed.
\item
An \textit{autoencoder} is a type of neural network trained to compress data into a lower-dimensional form (the “encoding”) and then to reconstruct the original data from that compressed version (the ``decodeing''). It learns to capture the most essential features of the input, making it useful for tasks like denoising, dimension reduction, or anomaly detection. Autoencoder has a lot in common with nonlinear principal components analysis and can have some of the look and feel of factor analysis if linearity is imposed. It has some features of machine learning but lacks others. The input data to autoencoder is unlabeled. For example, suppose for predictive policing, a plan is to enable police officers to identify emerging crime ``hot spots'' using their cell phones while on patrol. A full algorithm might be impractical for ``hand helds." An algorithm using many fewer predictors might be accurate enough and feasible. The small number of reconstituted predictors that are sufficient might be a product of autoencoder dimension reduction. 
\item
A \textit{generative adversarial network} (a GAN)  consists of two neural networks. One is a ``generator'' that creates synthetic data. The other is ``discriminator'' that tries to distinguish the real from the fake. They train together in competition, gradually improving until the generator produces highly realistic outputs. In a sense, the discriminator neural network helps to keep the generator neural network on task and honest. It can be seen as a more elaborate form of machine learning in which learning is improved through competition between algorithms.
\end{itemize}
\end{itemize}

\section{Yardsticks for AI Performance}

Probably the three most widely used AI performance metrics are accuracy, transparency, and fairness. Depending on the particular AI procedure, some are more salient than others. All three are not easily defined and can devolve into hand waving. 

Accuracy is complicated by a number of technical matters that depend on the choice of objective function. For categorical outcomes, such as forecasts of re-arrested, should false positives and false negatives be given the same weight? if not, what should their weights be? For numerical outcomes, such as forecasts of the numbers of times a police officer uses unnecessary force, should underestimates of the outcome be treated the same as overestimates of the outcome? How this is resolved affects which objective function is applied. 

Transparency can be a fuzzy concept at least until there is a satisfactory answer to the question ``transparent for whom?'' Large language models are not thoroughly understood by anyone, including those who invented them. Many very talented statisticians and computer scientists are working on the problem. Even the most simple AI procedures that are well understood by experts often will be poorly understood by most policy stakeholders, despite efforts to provide procedures that enable a look inside the metaphorical black box (Gilpin et al., 2018; Carvalho et al., 2019). Perhaps the easiest solution is to focus on the outcomes from an AI procedure, not on how the outcomes are produced. For example, if the concern is with fairness, hypothetical examples can be constructed that are similarly situated except for a protected class and see if the protected class membership makes a difference in the AI output. If unfairness is evident, it can be up to the experts to figure out what may be responsible. Further diagnostics can be employed (Molnar, 2025).

Fairness is often conflated with other concepts. For example, large disparities between sentence lengths for male and females offenders is not necessarily unfair. There can be legitimate reasons for the disparity and concerns about whether male offenders and females offenders are likely to be similarly situated. But in many discussions of fairness, a substantial disparity will automatically be called unfair. 

There also are many different kinds of unfairness that are often confused. For instance, ``demographic unfairness'' results when members of a particular protected class are overrepresented among those forecasted to be re-arrested compared to their representation in a relevant population. ``Forecasting accuracy unfairness'' occurs when forecasts from a risk algorithm are less accurate for one protected class compared to another. 

Perhaps more fundamentally, a risk algorithm that is fair in the treatment of groups is not necessarily fair in the treatment of individuals (Berk et al., 2023). Suppose at arraignment the same probability of re-arrest is estimated for protected group A and protected group B. That result might be achieved by inappropriately predicting a higher risk probability for a group A member some of the time, and inappropriately predicting a higher risk probability for a group B member some of the time. Individual fairness is sacrificed to achieve group fairness. 

Regardless of the details, there are two clear lessons. First, there likely to be tradeoffs between the performance criteria. For example, there are different kinds of fairness and you can't have them all (Kleinberg at al., 2017; Pleiss et al., 2017)). Correcting inequalities for AI outcomes, for instance, can introduce treatment inequalities. If an AI procedure is already optimized for accuracy, requiring fairness as well likely will reduce accuracy. Second, drawing on ideas for Alan Turing, the proper benchmark for AI performance is human performance at the same task. Indeed, one definition of AI requires that the AI equal or exceed human performance on the same task. There is growing evidence that this can be a pretty low bar (Berk et al., 2016; Berk, 2017). For instance, is the reasoning behind the sentencing decisions by judges really transparent? Are the release decisions by magistrates at arraignments based on accurate assessment of possible recidivism? Are the housing decisions by prison staff really racial fair? In each instance, AI likely can do better.
The same conclusion applies to legacy criminal justice tools. In risk assessment, for instance, an algorithm just needs to perform better than alternative risk assessment approaches such as the LSI-R, ORAS, or COMPAS. That too is a pretty low bar.
 
\section{Conclusions}

Looking ahead, one major conclusion is that AI is a moving target. Its algorithms and platforms rapidly will evolve.  A second major conclusion is that there will always be tradeoffs between different performance criteria. This occurs when performance is measured against perfection and remains when the more sensible benchmark of human performance is used. Third, the use of AI will only expand in criminal justice and elsewhere. It is not going to disappear. Fourth, whether from evil, ignorance or sloth, some misuse of AI will continue. But that is true of almost any technology. Finally, a variety of government regulations will be imposed at the federal level and perhaps by states. There will be, moreover, important differences between countries. At one extreme, might be the EU. At the other extreme, might be China. Content, enforceability, and reach will vary by jurisdiction. There will surely be litigation. At the very least, the regulatory environment will be very complicated. The content of this chapter is best viewed as a snapshot at one point in time.\footnote
{
Some good textbooks include: 
Reinforcement Learning -- Goldberg(1989); 
General Purpose Somewhat Dated  -- Bishop (2005); 
General Purpose Somewhat Dated -- Hastie et al. (2009);
Genetic Algorithms -- Eban and Smith (2015); 
General Purpose Recent -- Goodfellow et al. (2017);
Adversarial Machine Learning -- Joseph et al. (2019);
Transfer Learning --Yang et al (2020);  
All require at least two good statistics courses and a good computer science course. A course in linear algebra, and two courses in calculus are not mandatory but would help. More information is available on each book at Amazon. I don't know of any textbooks written at the level of this paper that are technically sound. But by all means look around.
}  
\clearpage

\theendnotes

\section*{References}
\begin{description}
\item
Bellovin, S.M. (2025) ``Computer Science and the Law'' \textit{Communications of the ACM} 5/21/2025. https://cacm.acm.org/opinion/computer-science-and-the-law/.
\item
Berk, R.A. {2017}
``An Impact Assessment of Machine Learning Risk Forecasts on Parole Board Decisions and Recidivism.'' \textit{Journal of Experimental Criminology} 13(2): 633 -- 655.
\item
Berk, R.A., Sorenson, S.B, and Barnes, G. (2016.)
``Forecasting Domestic Violence: A Machine Learning Approach to Help Inform Arraignment Decisions.'' \textit{Journal of Empirical Legal Studies} 13(1): 94 -- 115.
\item
 Berk, R.A., Kuchibhotla, A.K., and Tchetgen Tchetgen, E. (2023)
 ``Fair Risk Assessments.'' \textit{Annual Review of Statistics and Its Application} 10:  165 --187.
\item
Binet, A. and Simon, T. (1907) \textit{Les Infants Anormaux} (Paris, A. Colin).
\item
Bishop, C.M. (2005 \textit{Pattern Recognition and Machine Learning} Springer.
\item
Breiman, L. (2001) ``Statistical modeling: Two Cultures.'' \textit{Statistical Science} 16(3) : 199 -- 231.
\item
Brunton, S.L. and Kutz, N. (2022) \textit{Data-Driven Science and Engineering: Machine Learning, Dynamical Systems, and Control}, second edition. Cambridge Press. 
\item
Carvalho, D., Pereira, E., and Cardoso, J.J. (2019) ``Machine Learning Interpretability: A Survey on Methods and Metrics.'' \textit{Electronics} 8(8): 832; https://doi.org/10.3390/electronics8080832.
\item
Christian, J., (2023) ``Model Verification for Population Detection of Counterfeits.'' \textit{Journal of Forensic Sciences} 68(6): 2169 -- 2183.
\item
Coglianese, C., and Lehr, D. (2018) ``Transparency and Algorithmic Governance.''\textit{Administrive Law Review} 71: 1-- 56.
\item
Cooper, A.F., and Abrams, E. (2021)
``Emergent Unfairness in Algorithmic Fairness-Accuracy Trade-Off Research.''
\textit{Ethics, and Society} July: 46 -- 54.
\item
Dillavou, S., B. D. Beyer,B.D., Stern, M., Liu, A.J., Miskin M.Z., Durian, D.J.,  (2024) ``Machine Learning Without a Processor: Emergent Learning in a Nonlinear Analog Network."  \textit{Proceedings of The National Academy of Sciences} 121, e2319718121. 
\item
Dwork, C., Hardt, M., Pitassi, T., Reingold, O., and Zemel, R. (2011) ``Fairness Through Awareness.''
\textit{TCS '12: Proceedings of the 3rd Innovations in Theoretical Computer Science Conference}: 214 -- 226.
\item
Eben, A.E.  and Smith J.E. (2015) \textit{Introduction to Evolutionary Computing} second edition. Cambridge.
\item
Farayola, M.M., Tal, I.,  Connolly, R., Saber, T., and  Bendechache, (2023). ``Ethics and Trustworthiness of AI for Predicting the Risk of Recidivism: A Systematic Literature Review.'' \textit{Information} 14 (8) 426. https://doi.org/10.3390/info14080426.
\item
Fitzpartick, D. (2025) ``Geoffrey Hinton Predicts Human Extinction At The Hands Of AI. Here’s How To Stop It.'' \textit{Fprbes} January 2, 2025.
\item
geeksforgeeks (2024) ``AI Chips.'' 10/1/2024. Online at https://www.  \newline geeksforgeeks.org/artificial-intelligence/ai-chips-what-they-are- \newline and-why-they-matter/.
\item 
Gilpin, L.H., Bau, D., Yuan, B.Z., Bajwa, A., Specter, A., and Kagal, L.  (2018) ``Explaining Explanations: An Overview of Interpretability of Machine Learning.''  \textit{IEEE 5th International Conference on Data Science and Advanced Analytics} (DSAA), Turin, Italy: 80-89, doi: 10.1109/DSAA.2018.00018.
\item
Goldberg, D.E. (1989) \textit{Genetic Algorithms} Addison Wesley.
\item
Goldstein, M. (2002) \textit{The Complete Idiot's Guide to Weather}. Alpha Books.
\item
Goodfellow, I., Bengio, Y., and Courville, A. (2016) \textit{Deep Learning}. MIT Press.
\item
Hao, Z., Liu, S., Zhang, Y., Ying, C., Feng, Y., Su, H., and Zhu, J. (2022). ``Physics‑Informed Machine Learning: A Survey on Problems, Methods and Applications.'' \textit{arXiv preprint arXiv:2211.08064}.
\item
Hastie T, Tibshrani R, and Friedman J. (2009) \textit{Elements of Statistical Learning}, second edition Springer.
\item
Jones, C. (2021) ``Law Enforcement Use of Facial Recognition: Bias, Disparate Impacts on People of Color, and The Need for Federal Legislation.'' \textit{North Carolina Journal of Law \& Technology} 22(4): 778 -- 783.
\item
Imai, K., Jiang, Z., D James Greiner, J.D., Halen, R., and Shin, S. (2023) ``Experimental Evaluation of Algorithm-Assisted Human Decision-Making: Application to Pretrial Public Safety Assessment.'' \textit{Journal of the Royal Statistical Society Series A: Statistics in Society}186(2): 167 -- 189. https://doi.org/10.1093/jrsssa/qnad010.
\item
Johnson, M, (2025) \textit{Super AI} Google Books.
\item
Joseph, A.D., Nelson, B., Rubinstein, B.I.P., and Tygar, J.D. (2017) \textit{Adversarial Machine Learning} Cambridge.
\item
Kearns, M. and Roth, A. (2020) \textit{The Ethical Algorithm} Oxford Press.
\item
Kleinberg J, Mullainathan SM, and Raghavan M. (2017)  ``Inherent Tradeoffs in the Fair Determination of Risk Scores.'' \textit{Proc. 8th Conference on Innovations in Theoretical Computer Science} (ITCS).
\item
Kleinberg J, Himabindu L, Leskovec J, Ludwig J. and Sendhil M. (2018) ``Human Decisions and Machine Predictions.'' \textit{Quarterly Journal of Economics} 133 (1): 237 -- 293.
\item
Lane, S.M., and Houston, K.A. (2021) \textit{Understanding Eyewitness Memory} NYU Press.
\item
Mitchell S, Potash E, Barocas S, D'Amour A, and Lum K. (2021) ``Algorithmic Fairness, Choices, Assumptions and Definitions.'' \textit{Annual Review of Statistics and Its Applications} 8: 141 -- 163.
\item
Molnar, C. (2025) \textit{Interpretable Machine Learning: A  Guide for Making Black Box Models Explainable.} Learnpub, Google Books.
\item
Nielsen, M.A. (2015) \textit{Neural Networks and Deep Learning}. Determination Press.
\item
Neisser, U., Boodoo, G., Bouchard, T. J., Boykin, A. W., Brody, N., Ceci, S. J., ... and Urbina, S. (1996). ``Intelligence: Knowns and Unknowns.'' \textit{American Psychologist} 51(2), 77 -- 101. https://doi.org/10.1037/0003-066X.51.2.77.
\item
Pahune, S., Akhtar, Z., Mandapati, V., and Siddique, K. (2025) ``The Importance of AI Data Governance in Large Language Models.'' \textit{Big Data and Cognitive Computing}. 9: 147. https://doi.org/10.3390/bdcc9060147.
\item
Palmer, S. (2007). ``Lloyd, Edward (c.1648–1713).''  \textit{Oxford Dictionary of National Biography} Oxford University Press. doi:10.1093/ref:odnb/16829
\item
Petkauskas, V., and Japienyte, J. (2025) ```6 Billion Passwords Exposed in Record-Breaking Data Breach, Opening Access to Facebook, Google, Apple and Any Other Service imaginabl.'' \textit{Cybernews} 6/21/1925.
\item
Pleiss, G., Raghavan, M., Wu, F., Kleinberg, J., and Weinberger, K.Q. (2017) ``On Fairness and Calibration.'' In \textit{Advances in Neural Information Processing Systems}, eds. I. Guyon and U. Von Luxburg and S. Bengio and H. Wallach and R. Fergus and S. Vishwanathan and R. Garnett. \textit{Curran, Associates}.
\item
Saghiri, A. M., Vahidipour, S. M., Jabbarpour, M. R., Sookhak, M., and Forestiero, A. (2022). ``A Survey of Artificial Intelligence Challenges: Analyzing the Definitions, Relationships, and Evolutions.'' \textit{Applied Sciences}12(8), 4054. https://doi.org/10.3390/app12084054.
\item
Simon, A. (1956). ``Rational Choice and the Structure of the Environment.''  \textit{Psychological Review} 63(2): 129 -- 138.
\item
Stancati, M., Hinshow, D., Hagey, K and Glazer, E. (2025) ``Pople Leo Takes on AI as a Potential Threat to Humanity.'' \text{Wall Stree Journal} 6/17/2025.
\item
Stigler, S.M. (1990) \textit{The History of Statistics: The Measurement of Uncertainty before 1900} Belnap Press.
\item
Stigler, S.M. (2002) \textit{Statistics on the Table: The History of Statistical concepts and Methods} Harvard university Press.
\item
Stokel-Walker, C. (2025) ``Mark Zuckerberg's Superintelligence Gamble: Can Billions and Bold Hires Save Meta's AI Ambitions?'' \textit{Fast Company}. 6/11/2025.  https://www.fastcompany.com/91350445/meta-mark-zuckerberg-superintelligence-ai.
\item
Tierie, G. (1932) \textit{Cornelis Drebbel}. Amsterdam: HJ Paris.
\item
TRULEO Inc. (2025) 141 Traction St, Unit \#1040, Greenville, SC 29611
\item
Viswanathan, S.M, (2020) ``AI Chips: A New Semiconductor Era.'' \textit{International Journal of Advanced Research in Science, Engineering, and Technology} 7(8): 14683 -- 14694.
\item
G. Vrban\v{c}i\v{c} and V. Podgorelec, (2020) "Transfer Learning With Adaptive Fine-Tuning," \textit{IEEE Access.}" 8: 196197 -- 196211. doi: 10.1109/ACCESS.2020.3034343.
\item
Weiss, K., Khoshgoftaar, T.M., and Wang, D. (2016) ``A Survey of Transfer Learning.'' \textit{Journal of Big Data} 3(9) DOI 10.1186/s40537-016-0043-6.
\item
Wood, S.N. (2017) \textit{Generalized Additive Models: An Introduction with R}, second edition. Chapman \& Hall.
\item
Yang, Q., Zhang, Y., Dai, W., and Pan, S.J. (2020) \textit{Transfer learning} Cambridge.
\end{description}

\end{document}